\documentclass[10pt,conference]{IEEEtran}
\IEEEoverridecommandlockouts

\usepackage[normalem]{ulem}
\usepackage[utf8]{inputenc}
\usepackage{amsmath}  
\usepackage{graphicx}
\usepackage{comment}
\usepackage{multirow}
\usepackage{enumitem}
\usepackage{array}
\usepackage{tcolorbox}
\usepackage{xcolor}
\usepackage{balance}
\usepackage{mathtools}
\usepackage{xspace}
\usepackage{url}
\usepackage{eucal}
\usepackage{amsfonts}
\usepackage{color}
\usepackage{booktabs}
\usepackage{bbding}
\usepackage{float}

\usepackage{graphicx}
\usepackage{hyperref}
\usepackage{url}
\usepackage{multirow}
\usepackage{comment}
\usepackage{colortbl}
\usepackage{graphicx, subfigure}

\usepackage{xcolor}
\usepackage{enumitem}
\usepackage{amsmath,amssymb}

\def\BibTeX{{\rm B\kern-.05em{\sc i\kern-.025em b}\kern-.08em
    T\kern-.1667em\lower.7ex\hbox{E}\kern-.125emX}}
    
\begin{document}
\title{Generation-based Code Review Automation: \\ How Far Are We?}

\makeatletter
\newcommand{\linebreakand}{%
  \end{@IEEEauthorhalign}
  \hfill\mbox{}\par
  \mbox{}\hfill\begin{@IEEEauthorhalign}
}
\makeatother

\author{\IEEEauthorblockN{Xin Zhou}
\IEEEauthorblockA{\textit{Singapore Management University} \\
Singapore \\
xinzhou.2020@phdcs.smu.edu.sg}
\and
\IEEEauthorblockN{Kisub Kim\textsuperscript{*}\thanks{*Corresponding author. Email: kisubkim@smu.edu.sg}}
\IEEEauthorblockA{\textit{Singapore Management University} \\
Singapore \\
kisubkim@smu.edu.sg}
\and
\IEEEauthorblockN{Bowen Xu}
\IEEEauthorblockA{\textit{Singapore Management University} \\
Singapore \\
bowenxu@smu.edu.sg}
\linebreakand
\IEEEauthorblockN{DongGyun Han}
\IEEEauthorblockA{\textit{Royal Holloway, University of London} \\
UK \\
donggyun.han@rhul.ac.uk}
\and
\IEEEauthorblockN{Junda He}
\IEEEauthorblockA{\textit{Singapore Management University} \\
Singapore \\
jundahe@smu.edu.sg}
\and
\IEEEauthorblockN{David Lo}
\IEEEauthorblockA{\textit{Singapore Management University} \\
Singapore \\
davidlo@smu.edu.sg}
}

\maketitle

\begin{abstract}

Code review is an effective software quality assurance activity; however, it is labor-intensive and time-consuming. 
Thus, a number of generation-based automatic code review (ACR) approaches have been proposed recently, which leverage deep learning techniques to automate various activities in the code review process (e.g., code revision generation and review comment generation).

We find the previous works carry three main limitations. First, the ACR approaches have been shown to be beneficial in each work, but those methods are not comprehensively compared with each other to show their superiority over their peer ACR approaches. 
Second, general-purpose pre-trained models such as CodeT5 are proven to be effective in a wide range of Software Engineering (SE) tasks. 
However, no prior work has investigated the effectiveness of these models in ACR tasks yet. 
Third, prior works heavily rely on the Exact Match (EM) metric which only focuses on the perfect predictions and ignores the positive progress made by incomplete answers.
To fill such a research gap, we conduct a comprehensive study by comparing the effectiveness of recent ACR tools as well as the general-purpose pre-trained models. 
The results show that a general-purpose pre-trained model CodeT5 can outperform other models in most cases. 
Specifically, CodeT5 outperforms the prior state-of-the-art by 13.4\%--38.9\% in two code revision generation tasks. 
In addition, we introduce a new metric namely Edit Progress (EP) to quantify the partial progress made by ACR tools.
The results show that the rankings of models for each task could be changed according to whether EM or EP is being
utilized.
Lastly, we derive several insightful lessons from the experimental results and reveal future research directions for generation-based code review automation.

\end{abstract}

\section{Introduction}
\label{sec:intro}

Modern software development involves numerous software quality assurance activities, such as defect management~\cite{wang2016automatically,saha2013improving,wang2014version}, testing~\cite{myers2011art,garousi2016and}, and code review~\cite{bosu2016process,bacchelli2013expectations}, to ensure the quality of software. Code review specifically requires reviewers to assess whether the source code written by authors satisfies both functional (e.g., compilation and testing) and non-functional (e.g., code readability) requirements. 
Many studies~\cite{rigby2013convergent,mcintosh2016empirical,bavota2015four,morales2015code, mcintosh2014impact,sadowski2018modern,beller2014modern} have shown the outstanding effectiveness of the code review process in removing defects and improving maintainability and readability. 
In addition, code review also helps to share programming knowledge among developers~\cite{bacchelli2013expectations, sadowski2018modern}.

The benefits of code reviews have been broadly recognized and it is widely adopted in both proprietary and open-source software projects. 
However, the benefits come with extensive manual costs in reviewing. 
In the modern software development process, a large number of code changes are required to be reviewed per month. 
For example, Microsoft Bing and Linux projects require 3,000 and 500 reviews, respectively, per month~\cite{rigby2013convergent,rigby2014peer}.
In addition, studies~\cite{ko2006exploratory,minelli2015know,zelkowitz1979principles,xia2017measuring} show that program comprehension (one of the major work for code reviewers) takes up as much as half of a developer’s time, especially when the code quality is low~\cite{xia2017measuring}. 
To reduce the developers' burden, researchers have created techniques~\cite{tufano2021towards,thongtanunam2022autotransform,tufano2022using} to automate code review activities by leveraging deep learning algorithms.
Particularly, they focus on revising the submitted code to address the possible flaws in the code (i.e., Code Revision Before Review)~\cite{tufano2021towards,thongtanunam2022autotransform,tufano2022using}, writing review comments based on the submitted code (i.e., Review Comment Generation)~\cite{tufano2022using}, and revising the submitted code based on the comments written by reviewers (i.e., Code Revision After Review)~\cite{tufano2021towards,tufano2022using}.

Recent approaches~\cite{tufano2021towards,thongtanunam2022autotransform,tufano2022using} targeting the above code review tasks push code review automation to new heights.
Tufano et al.~\cite{tufano2021towards} proposed an approach namely Trans-Review based on Transformer~\cite{transformer} to complete the code revision generation tasks. 
They first abstracted the source code by adopting the src2abs tool~\cite{Tufano-Learning-CodeChanges} to reduce the vocabulary size. Then they trained a sequence-to-sequence Transformer on the abstracted code. 
Thongtanunam et al.~\cite{thongtanunam2022autotransform} proposed AutoTransform, which leverages the Byte-Pair Encoding (BPE)~\cite{BPE} tokenizer and a sequence-to-sequence Transformer to complete the code revision generation task. They highlighted the problem that the revised version of the code might have new identifiers/literals (i.e., new tokens), and it is challenging to generate those new tokens. The BPE approach could help to mitigate such a problem.
Tufano et al.~\cite{tufano2022using} leveraged the Text-To-Text Transfer Transformer (T5) architecture~\cite{T5} to develop a pre-trained model for code review (i.e., T5-Review) based on 1.4 million code snippets and texts from Stack Overflow\footnote{https://stackoverflow.com/} and CodeSearchNet~\cite{codesearchnet}. 
The results showed that the pre-trained T5-Review led to a significant improvement over the non-pre-trained Transformer models in three generation-based code review tasks.

Although ACR methods have been proven to be beneficial, 
these methods are not compared against each other which may confuse the practitioners about which one is suitable for a specific task.
In addition, general-purpose pre-trained models for code like CodeBERT~\cite{codebert}, GraphCodeBERT~\cite{GraphCodeBERT}, and CodeT5~\cite{codeT5} have been proven to be effective in a wide range of downstream tasks~\cite{zhou2021assessing,lu2021codexglue} in the field of software engineering such as code search and code summarization. 
However, the performance of these models on ACR remains unknown.

\begin{figure}
	\centering
	\includegraphics[width=1\columnwidth]{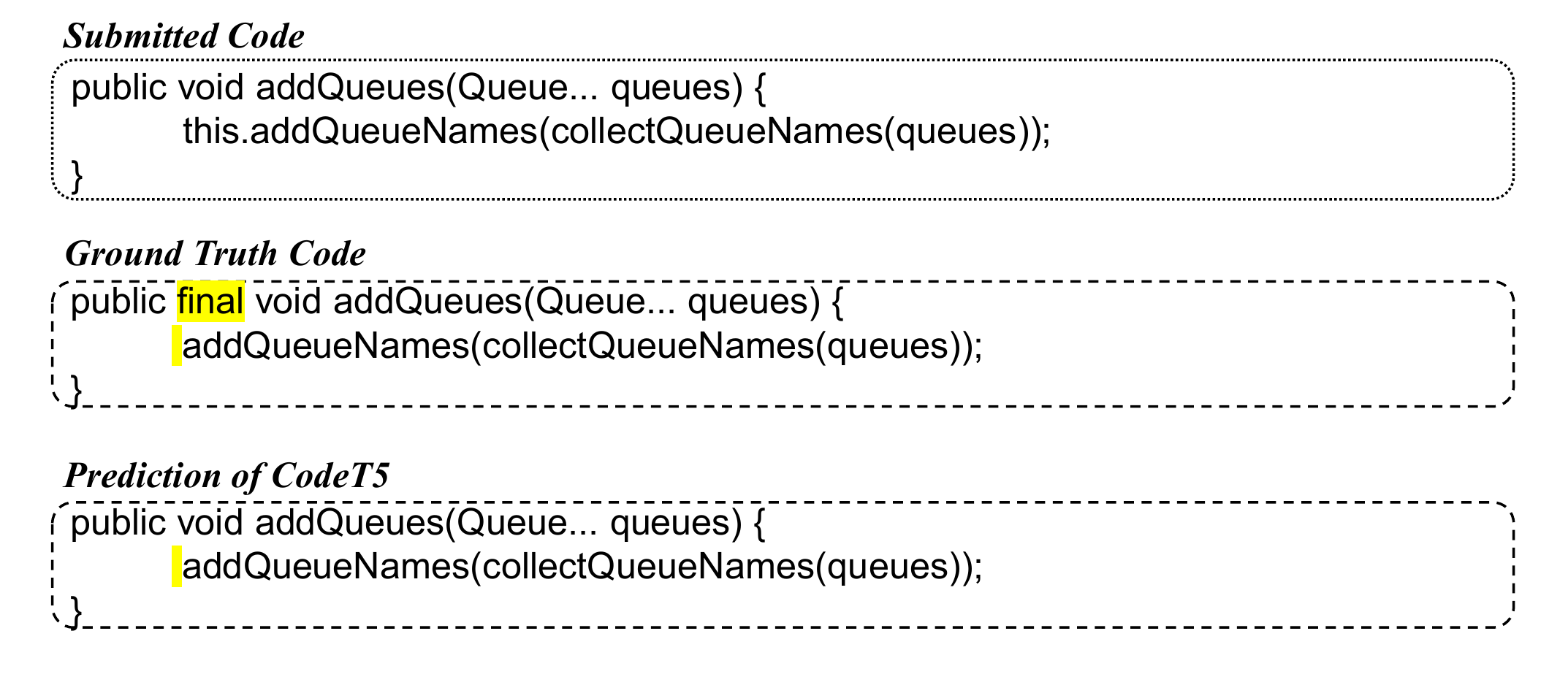}
 \vspace{-0.5cm}
	\caption{A model prediction with a positive contribution though not perfect. Ground truth code that is accepted by reviewers removes the ``this'' keyword and adds the ``final'' keyword (as highlighted). The model prediction of CodeT5 only removes the ``this'' keyword but still improves the initial submitted code.}
	\label{fig:ep-example}
\end{figure}

Existing ACR works heavily rely on a strict evaluation metric called Exact Match (EM): a revised code is considered as correct only if it is identical to the ground truth code.
Although Exact Match is useful, it is very strict and only focuses on the number of perfect predictions.
Prior studies~\cite{not-replace,kochhar2015understanding,kochhar2016practitioners} indicated that the developers did not blindly trust automated tools and their recommendations. For instance, Winter et al.~\cite{not-replace} surveyed 386 software developers about their attitudes toward automatic program repairs (APR) tools which are similar to ACR tools. 47.0\% of responses highlighted that human decisions are needed when applying recommendations from automated tools and 23.0\% of responses do not trust automatic tools at all.
In addition, the perfect prediction ratios of ACR tools are still low (i.e., 1.2\%--23.2\%). Thus, 
we believe that the current ACR tools are targeting to provide high-quality drafts for code reviewers rather than fully replacing them. 
In this sense, instead of only focusing on the number of perfect predictions (EM), we need to care more about the improvements made in the predictions compared to the initial submitted code on the average level.
Figure~\ref{fig:ep-example} shows an example that CodeT5 generates a better draft code based on the submitted code by removing ``\texttt{this.}’’. Though the generated code cannot meet all requirements of reviewers, it still makes positive progress, which should be taken into account when evaluating the effectiveness of a model.
To quantify the progress made in predictions, we introduce a new evaluation metric namely Edit Progress (EP), which measures the reduction ratio of edits needed from the submitted code to the ground truth.

To fill the gaps, we conduct a large-scale comprehensive experiment by comparing the effectiveness of recent ACR tools as well as the general-purpose pre-trained models on a unified benchmark. 
Our goal is to figure out how far we have reached in generation-based code review tasks and highlight the challenges and potential research directions for better code review automation.

In summary, our study covers three state-of-the-art ACR tools (i.e., Trans-Review~\cite{tufano2021towards}, AutoTransform ~\cite{thongtanunam2022autotransform}, and T5-Review~\cite{tufano2022using}) and three general-purpose pre-trained models (i.e., CodeBERT~\cite{codebert}, GraphCodeBERT~\cite{GraphCodeBERT}, and CodeT5~\cite{codeT5}) on three code review downstream tasks: 
Code Revision Before Review (i.e., revising the submitted code to address the possible flaws), Code Revision After Review (i.e., revising the submitted code based on the comments written by reviewers), and Review Comment Generation (i.e., writing review comments pointing out problems in the submitted code).

Our experimental results reveal the following key findings:

\begin{enumerate}[leftmargin=*]
\item \textit{A pre-trained encoder-decoder model, CodeT5, achieves state-of-the-art performance and outperforms all existing ACR techniques in most cases.}  
We carry out a comprehensive evaluation and discover that 1) T5-Review is the best model among existing ACR approaches in terms of the Exact Match metric; 2) CodeT5 can outperform all existing ACR approaches by a large margin for most cases.

\item \textit{Stack Overflow data is helpful to the Review Comment Generation task.}
As more pre-training corpus written in natural language are provided in Stack Overflow data, T5-Review which is pre-trained on Stack Overflow data has significantly outperformed all other methods by a large margin.

\item \textit{Partial progress also matters.}
We observe that a model that can generate the most perfect predictions may not able to generate the most close-to-perfect predictions, indicating that there may exist much noise on the failure cases of the approaches that can generate many perfect predictions.
In addition, the rankings of approaches in EM and EP can be different. This indicates that we may miss tools that can contribute the most positive progress if it ranks not very high in EM. Thus, measuring partial progress is also important for a comprehensive understanding of the ACR tools' effectiveness. 

\end{enumerate}

Overall, our empirical investigation sheds light on the strength and weaknesses of each ACR approach as well as the future research directions for generation-based code review automation.
We summarize our contributions as follows: 
\begin{enumerate}[leftmargin=*]
\item We conduct a comprehensive investigation on the effectiveness of existing ACR models as well as pre-trained models for code in three downstream code review tasks.
\item We introduce a novel evaluation metric to quantify the progress made by ACR tools in generating revised code.
\item We derive several insightful lessons from the experimental results and reveal the future direction of the technical design for code review tasks.

\end{enumerate}

\section{Background}
\label{sec:background}

\begin{figure}
	\centering
	\includegraphics[width=1\columnwidth]{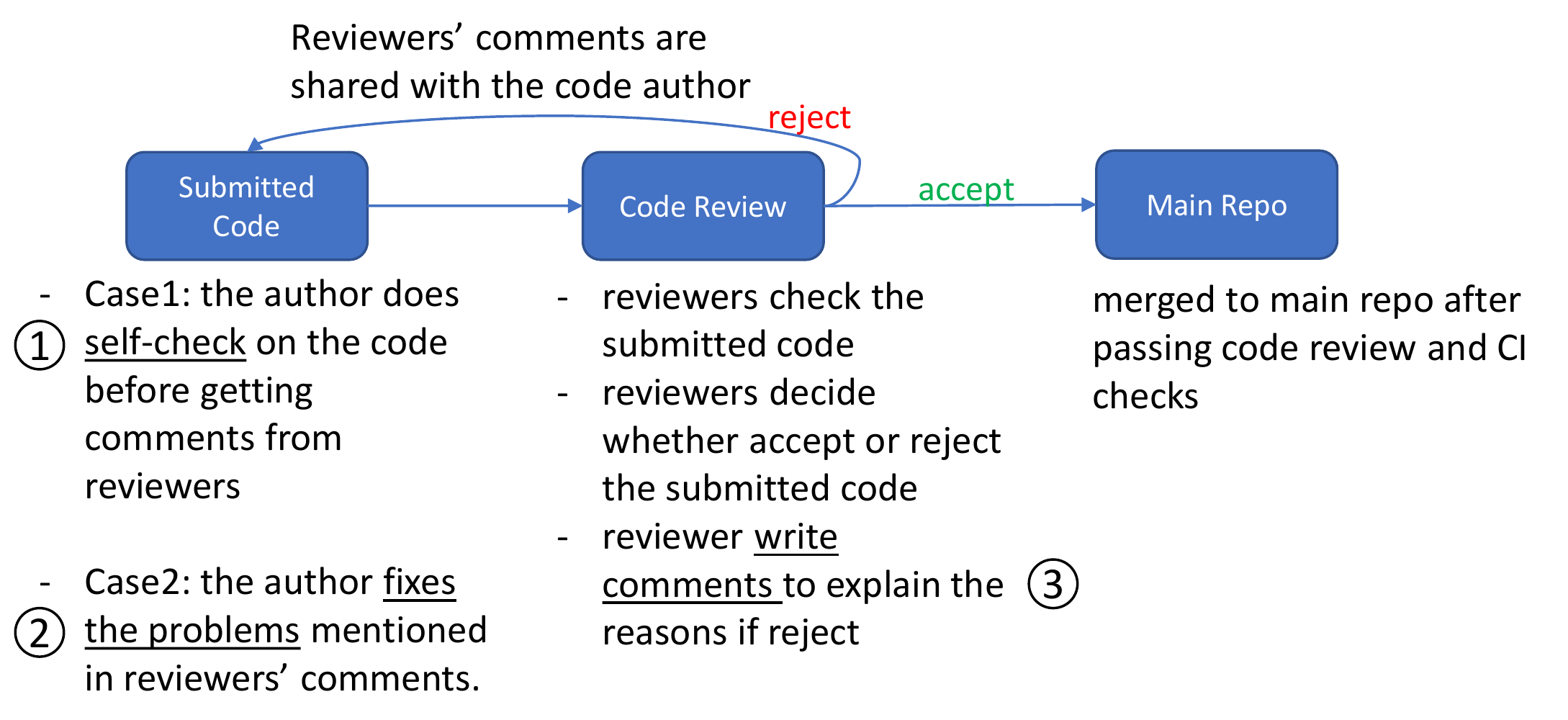}
	\caption{A brief introduction about the code review processes. The \textcircled{1}, \textcircled{2}, and \textcircled{3} refer to the three code review activities that generation-based automatic code review tasks automate.}
	\label{fig:review_process}
\end{figure}

In this section, we introduce the code review processes and formulate the three automatic code review tasks. 
Then we briefly summarize the automatic code review tools and general-purpose pre-trained code models under investigation.

\subsection{Code Review Processes}
\label{sec:activity}
Figure~\ref{fig:review_process} briefly shows how code reviews are performed in software development. Code authors first check the correctness of the code by themselves and then submit the code for review. Code reviewers check the submitted code and decide whether to accept/reject to merge the code into the main repository. If the submitted code is rejected, reviewers write comments explaining what and how to improve the code.
Then the code authors address those comments and re-submit for review again.

\subsection{Generation-based Code Review Tasks}
\label{sec:tasks}
In this subsection, we describe the generation-based code review downstream tasks. The considered tasks in our study are all proposed to automate the subset of activities in code review processes. We follow the prior studies~\cite{tufano2021towards, thongtanunam2022autotransform,tufano2022using} to formulate each of those tasks.

\vspace{0.1cm}
\noindent\textbf{Code Revision Before Review (CRB)} mainly aims to help the code authors who make the code changes to improve and commit them to the repositories.
It could help the code authors to address some simple flaws and thus improve the quality of the code \textit{before submitting it for review}.
It automates the code review activity \textcircled{1} in Figure~\ref{fig:review_process}.
At the same time, it also reduces the burden of reviewers on pointing out simple and repeated flaws. 
This task is formulated as a sequence-to-sequence task: $f(Code)\xrightarrow[]\, Revised \, Code $ where $f$ is the DL model and ``$Revised \, Code$'' is the ground truth (i.e., the revised code that is accepted by code reviewers).

\vspace{0.1cm}
\noindent\textbf{Code Revision After Review (CRA)} 
aims at supporting code authors by addressing review comments
which automate the code review activity \textcircled{2} in Figure~\ref{fig:review_process}).
The techniques belonging to this task can boost the speed of the revision process as they generate the initial version.
This task is formulated as a sequence-to-sequence task: $f(Code, Comment)\xrightarrow[] \, Revised \, Code $ where $f$ is the DL model and ``$Revised \, Code$'' is the ground truth code accepted by reviewers.
For this task, there have two inputs (i.e., the submitted code and the comments) and a single output (i.e., the revised code).

\vspace{0.1cm}
\noindent\textbf{Review Comment Generation (RCG)} is designed for the reviewers.
The models in this task draft the initial review comments such that reviewers can save time by revising the written draft, which automates the code review activity \textcircled{3} in Figure~\ref{fig:review_process}.
Review Comment Generation is also formulated as a sequence-to-sequence task: $f(Code)\xrightarrow[]\, Comment$ where $f$ is the DL model.

\begin{figure}
	\centering
	\includegraphics[width=1\columnwidth]{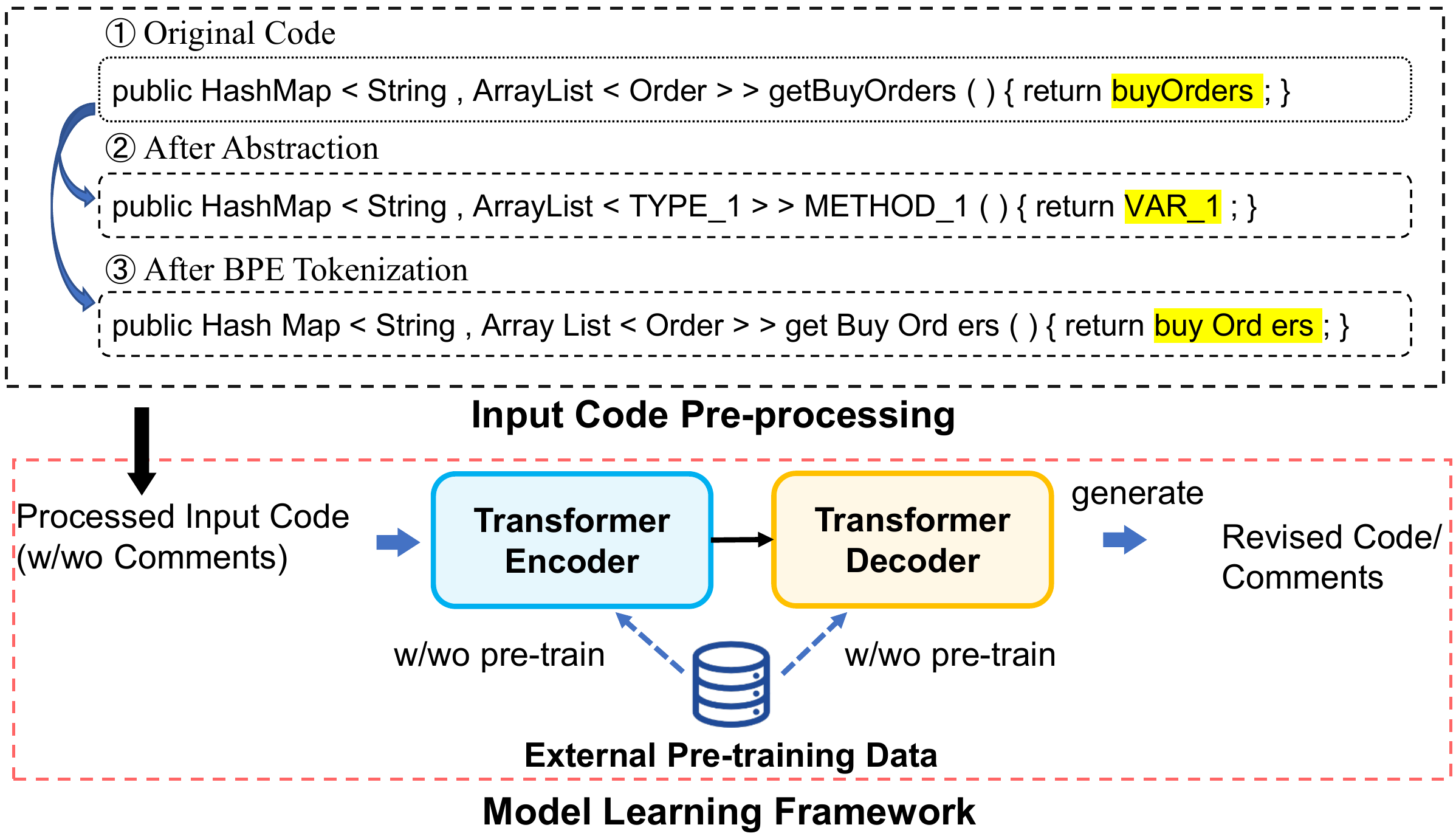}
 \vspace{-0.5cm}
	\caption{A Unified Model/Task Framework. All studied models can be fitted in this unified framework by choosing the pre-precessing strategy and Transformer models with/without pre-training. All ACR tasks can be fitted into this framework by choosing the corresponding input-output pairs. Note that ``w" refers to ``with" and ``wo" refers to ``without". The parts highlighted in yellow are examples of different pre-processing strategies.}
	\label{fig:models}
\end{figure}

\subsection{Automatic Code Review  Tools}
\label{sec:models}

All considered tools are built based on a popular Deep Learning model architecture called Transformer~\cite{transformer}, which mitigates the long-distance dependency issue of CNN-based models~\cite{linsley2018learning} and the long-time computation issue of RNN-based models~\cite{transformer}.
A Transformer model typically has two parts: an encoder to understand the semantics of the input data and a decoder to generate the final results (e.g., description, translation, revision, etc.) 
All studied models adopt the Transformer model (both the encoder and decoder). We highlight the novel designs of each studied model in the following.

\vspace{0.1cm} 
\noindent\textbf{Trans-Review}~\cite{tufano2021towards} are proposed to complete the Code Revision Before Review task (CRB) and the Code Revision After Review (CRA).
A large vocabulary size will hinder the learning process and lead to inaccurate generation~\cite{hellendoorn2017deep}. To reduce the vocabulary size, Trans-Review abstracted the source code by adopting the src2abs~\cite{Tufano-Learning-CodeChanges}, which replaces actual identifiers/literals in code with reusable IDs.
Specifically, for both the submitted and revised versions of code, it replaces identifiers (e.g., types and names of source code elements) and literals (e.g., string values) with an ID that is allowed to be reused across different code snippets. 
For instance, as shown in Figure~\ref{fig:models}, the \textit{first} variable appearing in each code will always be replaced with an ID of ``1'' and a string ``VAR'' to indicate it is a variable (i.e., VAR\underline{\hspace{0.5em}}1).

\vspace{0.1cm}
\noindent\textbf{AutoTransform}~\cite{thongtanunam2022autotransform} is only designed to complete the Code Revision Before Review task (CRB).
Unlike Trans-Review, it applies a Byte-Pair Encoding (BPE)~\cite{BPE} approach to reduce the size of the vocabulary.
BPE is a tokenization approach that splits a code token into a sequence of frequently occurring subtokens. For instance, a code token ``buyOrders" in Figure~\ref{fig:models} will be split into several subtokens ``buy'', ``Ord'', and ``ers''. 
This strategy brings benefits to efficiency as subtokens are not only occurred more frequently than the regular code tokens, but they also allow to reduce the size of the vocabulary.
Empirical evidence showed that BPE could effectively address the large vocabulary size  problem~\cite{karampatsis2020big}.

\vspace{0.1cm}
\noindent\textbf{T5-Review}~\cite{tufano2022using} leveraged the Text-To-Text Transfer Transformer (T5) architecture~\cite{T5}. 
T5 is a pre-training paradigm that corrupts the input data (e.g., randomly masking/dropping/inserting tokens) before feeding into the Transformer encoder and reconstructing the original input data at the Transformer decoder. Those pre-training tasks are called denoising objectives.
Tufano et al. pre-trained a code review version of T5 (denoted T5-Review) on 1.4 million code snippets and texts from Stack Overflow and the CodeSearchNet dataset~\cite{codesearchnet}. T5-Review consists of a 6-layer Transformer encoder and a 6-layer Transformer decoder. To reduce the vocabulary size, T5-Review leverages BPE tokenization~\cite{BPE}.

\subsection{General-purpose Pre-trained Models for Code}
The following techniques are the popular pre-trained models that are trained on source code and texts.

\vspace{0.1cm}
\noindent\textbf{CodeBERT}~\cite{codebert} is a SE knowledge-enriched bi-modal pre-trained model, which is capable of modeling both natural languages (NL) and programming languages (PL). 
CodeBERT leverages a 12-layer Transformer encoder as the model and pre-trained the encoder with the NL-PL data pairs from the CodeSearchNet dataset~\cite{codesearchnet}.
It adopts two pre-training objectives jointly: Masked Language Modeling (MLM)~\cite{bert} and Replaced Token Detection (RTD)~\cite{electra}. 
The eventual loss function for CodeBERT is formulated below:
\begin{equation}
    \underset{\theta}{\text{min}}( 
    \mathcal{L}_{MLM}(\theta)+\mathcal{L}_{RTD}(\theta))
\end{equation} 
where $\theta$ denotes the model parameters of the 12-layer Transformer encoder.
CodeBERT has shown great effectiveness in a range of SE downstream tasks such as code search and code summarization~\cite{codebert,zhou2021assessing,lu2021codexglue}.

\vspace{0.1cm}
\noindent\textbf{GraphCodeBERT}~\cite{GraphCodeBERT} brings the inherent structure of code into consideration during source code modeling. 
To learn source code representation, GraphCodeBERT leverages three input components, which are Programming Language, Natural Language, and Data Flow Graph.
GraphCodeBERT introduces two structure-aware pre-training tasks (i.e., Edge Prediction and Node Alignment) aside from the MLM prediction task.

\vspace{0.1cm}
\noindent\textbf{CodeT5}~\cite{codeT5}
is another pre-trained model whose Transformer encoder and decoder are both pre-trained. CodeT5 is capable of establishing complex mappings from sequence to sequence, which is suitable for generation-based tasks. CodeT5 is developed based on the general T5 architecture and is pre-trained not only with the original pre-training tasks (i.e., the denoising objectives) of T5 but also a set of newly proposed pre-training tasks (e.g., Masked Identifier Prediction and Identifier Tagging). Specifically, CodeT5 is enhanced by making it aware of the type information of identifiers.

\section{Experimental Design}
\label{sec:settings}
In this section, we introduce our research questions and describe the corresponding experimental settings.

\subsection{Research Questions}
\textit{\textbf{RQ1. Which is the best-performing ACR tool for each targeted code review activity?}}
Recent ACR tools have been demonstrated to be effective in code review tasks.  
However, a comparative study is missing, which makes it confusing to rank the existing techniques. 
Addressing this research question may support practitioners in selecting the most suitable technique for each downstream task, and the revealed state-of-the-art that can be considered for the baseline of future research.
In this RQ, we conduct a revisiting evaluation of existing ACR tools on the same set of available datasets.

\textit{\textbf{RQ2. Can popular pre-trained models outperform the existing ACR models?}}
General-purpose pre-trained code models have shown significant generalizability and demonstrated promising performance in a broad range of SE tasks~\cite{codebert,codeT5,lu2021codexglue,zhou2021assessing} with datasets of varying sizes and properties.
However, prior ACR tools have not considered these models either as baselines or building blocks of the ACR tools. 
We believe that discovering the effectiveness of such models on code review tasks would contribute to the research community, and they can also be the baseline for future works.
The subject models are listed as follows: (1) CodeBERT~\cite{codebert}, GraphCodeBERT~\cite{GraphCodeBERT}, and CodeT5~\cite{codeT5}.
To answer this research question, we take popular pre-trained models into consideration and compare them with existing ACR models.

\textit{\textbf{RQ3. Comparing the generated code with the human-written code, how far apart are they?}}
The prior ACR works only use a strict evaluation metric (i.e., Exact Match) to analyze the effectiveness of their models: a revised code is considered as a correct generation only if it is identical to the ground truth. Otherwise, no matter if the difference is just a single token or they are different, the generated code is equally considered wrong. 
Thus, we may miss suggestions that are very close to ground truths that are still partially useful (as shown in the example in Figure~\ref{fig:ep-example}). 
It may be interesting to have an overall idea of how close those generated codes are compared to the ground truths because the current automation of the code review process does not only aim to replace the practitioners but also provide suggestions/drafts for developers to work on~\cite{not-replace}.
To answer this research question, we introduce a new evaluation metric, Edit Progress, which quantifies the progress achieved by the generated code from the problematic code toward the ground truth.
Edit Progress is firstly used in the text editing task to measure the progress~\cite{elgohary-etal-2021-nl}. 
Note that this metric can only be applied in code revision generation tasks because Edit Progress needs an initial input and the revised version to compute whether the revised one improves or degrades the initial one. However, in the Review Comment Generation task, there is no initial comment.

\subsection{Experimental Settings}

\subsubsection{\textbf{Datasets}}
We include three different datasets used in evaluating the latest ACR tools, namely $\text{Trans-Review}_\text{data}$, $\text{AutoTransform}_\text{data}$, and $\text{T5-Review}_\text{data}$.
Tufano et al.~\cite{tufano2021towards} collected the $\text{Trans-Review}_\text{data}$ from projects in Gerrit and GitHub.
They filtered out noisy comments and comments that are longer than 100 tokens. Besides, they removed the review data where the revised code has new tokens not shown in the initial submitted code.
Thongtanunam et al.~\cite{thongtanunam2022autotransform} used the $\text{AutoTransform}_\text{data}$ that is collected from three Gerrit code review repositories: Android\footnote{\url{https://android-review.googlesource.com/.}}, Google\footnote{\url{https://gerrit-review.
googlesource.com/}}, and Ovirt\footnote{\url{https://gerrit.ovirt.org/}}. Please note that $\text{AutoTransform}_\text{data}$ only has the submitted codes and the revised codes but does not have the corresponding review comments.
Tufano et al~\cite{tufano2022using} collected $\text{T5-Review}_\text{data}$ from Java open-source projects from GitHub which have at least 50 pull requests, 10 contributors, 10 stars, and are not forks. They filtered out noisy comments, non-English comments, and duplicate comments. 

In this paper, for the Code Revision Before Review (CRB), we employ $\text{Trans-Review}_\text{data}$, $\text{AutoTransform}_\text{data}$, and $\text{T5-Review}_\text{data}$.
For the Code Revision After Review (CRA) and Review Comment Generation (RCG), we leverage $\text{Trans-Review}_\text{data}$ and $\text{T5-Review}_\text{data}$. We do not use $\text{AutoTransform}_\text{data}$ because it does not contain the reviewers' comments given to submitted codes.
The statistics of the used datasets are summarized in Table~\ref{tab:datasets}.

\begin{table}[]
\centering
\caption{Statistics of Studied Datasets.}
\label{tab:datasets}
\begin{tabular}{@{}lrrrr@{}}
\toprule
\textbf{Dataset Statistics}                 & \multicolumn{1}{c}{\textbf{\#Train}} & \multicolumn{1}{c}{\textbf{\#Valid}} & \multicolumn{1}{c}{\textbf{\#Test}} & \multicolumn{1}{c}{\textbf{\begin{tabular}[c]{@{}c@{}}\#New\\ Tokens\end{tabular}}} \\ \midrule
\multicolumn{1}{l|}{\textbf{Trans-Review}}  & \multicolumn{1}{r|}{13,756}          & \multicolumn{1}{r|}{1,719}           & \multicolumn{1}{r|}{1,719}          & 0\%                                                                                 \\ \midrule
\multicolumn{1}{l|}{\textbf{AutoTransform}} & \multicolumn{1}{r|}{118,039}         & \multicolumn{1}{r|}{14,750}          & \multicolumn{1}{r|}{14,750}         & 82.9\%                                                                              \\ \midrule
\multicolumn{1}{l|}{\textbf{T5-Review}}     & \multicolumn{1}{r|}{134,239}         & \multicolumn{1}{r|}{16,780}          & \multicolumn{1}{r|}{16,780}         & 89.6\%                                                                              \\ \bottomrule
\end{tabular}
\begin{flushleft}
{*Note that the last column shows the ratio of instances that the ground truth revised code contains new tokens that are not shown in the initial submitted code.}
\end{flushleft}
\end{table}

\subsubsection{\textbf{Evaluation Metrics}}

Following previous literature~\cite{tufano2021towards, thongtanunam2022autotransform, tufano2022using}, we report the performance of each investigated model using the Exact Match metric (RQ1 and RQ2). We additionally report Edit Progress to measure how close generated codes are compared to the ground truths (RQ3).

\vspace{0.1cm}
\noindent\textbf{Exact Match (EM)} is a strict metric. 
Specifically, 
for each input-output pair, if the tokens of the model's prediction (generated by the model given the input) exactly match the tokens of the ground truth (output), EM = 1; otherwise, EM = 0. 
With just a single token difference between the generated code and the ground truth, the EM score of the generated code is 0.

\vspace{0.1cm}
\noindent\textbf{Edit Progress (EP)} is a novel metric to measure the progress made by the generated code on the path from the erroneous code to the clean code.
Recently, Dibia et al.~\cite{dibia2022aligning} highlighted the significance of measuring the effort to correct the generated code. They suggested that an edit distance-based metric~\cite{svyatkovskiy2020intellicode} could serve as an important proxy for estimating the effort required to edit and fix generated code.
As an edit distance-based metric, the EP metric may offer a better estimation of the user effort needed to edit code generations in ACR tasks than EM.
In automatic code review tasks,  each data instance in the test set contains an initial submitted code  $\widetilde{C}$ and a ground truth code $\bar{C}$ that is revised by developers and accepted by code reviewers. Given the initial submitted code $\widetilde{C}$, each ACR model generates a revised code $\hat{C}$. The Exact Match (EM) is a strict metric as EM = 1 only when $\hat{C}$ is identical to $\bar{C}$.
While useful, Exact Match also has limitations. It only expects models to be able to fully correct/revise an erroneous code into the correct code.  While in some cases, models are still able to make progress by reducing the number of errors in the submitted code, even though the generated code revisions $\hat{C}$ are not identical to the ground truth code revisions $\bar{C}$.
To measure the progress made on the path from the erroneous code to the clean code, we adopt the Edit Progress~\cite{elgohary-etal-2021-nl} metric in this work. 
Specifically, for a data instance $\langle \widetilde{C}, \bar{C} \rangle$, the Edit Progress (EP) scores of ACR models are computed as the relative edit reduction between the initial code $\widetilde{C}$ and the ground truth code $\bar{C}$ by predicting a revision $\hat{C}$. The metric is formulated in the following equation:

\begin{equation}
    Progress = \frac{|D_{\widetilde{C} \xrightarrow[]{} \bar{C}}|  -  |D_{\hat{C} \xrightarrow[]{} \bar{C}}|}{|D_{\widetilde{C} \xrightarrow[]{} \bar{C}}|}
\end{equation}
where $D$ is the edit distance~\cite{levenshtein1966binary} (i.e., the minimum number of operations required to transform one sequence into the other).
$|D_{A \xrightarrow[]{} B}|$ refers to the edit distance between a string A to a string B.
An example of how to calculate the EP score is shown in Figure~\ref{fig:edit_progress}. In the example, to turn the submitted code $\widetilde{C}$ into the corresponding ground truth code $\bar{C}$, there are four token-level edit actions: 
1) replace ``hypot'' with ``cosh'';
2) delete ``,'';
3) delete ``double'';
4) delete ``y''.
If a model generates a prediction like the first prediction in Figure~\ref{fig:edit_progress}, there are three deletions left. Thus, the prediction makes 25\% progress on the way from the submitted code to the ground truth revised code.
Similarly, the second prediction in Figure~\ref{fig:edit_progress} makes 75\% progress (reduce 75\% edits needed from the submitted code to the ground truth code).
A perfect prediction $\hat{C}$ that can fully correct all errors in the submitted code $\widetilde{C}$ would achieve a 100\% Edit Progress score.
A model prediction $\hat{C}$ can have negative progress if it generates more errors than correct edits. We calculate and report the average EP on all test data instances for each model.

\begin{figure}
	\centering
	\includegraphics[width=\columnwidth]{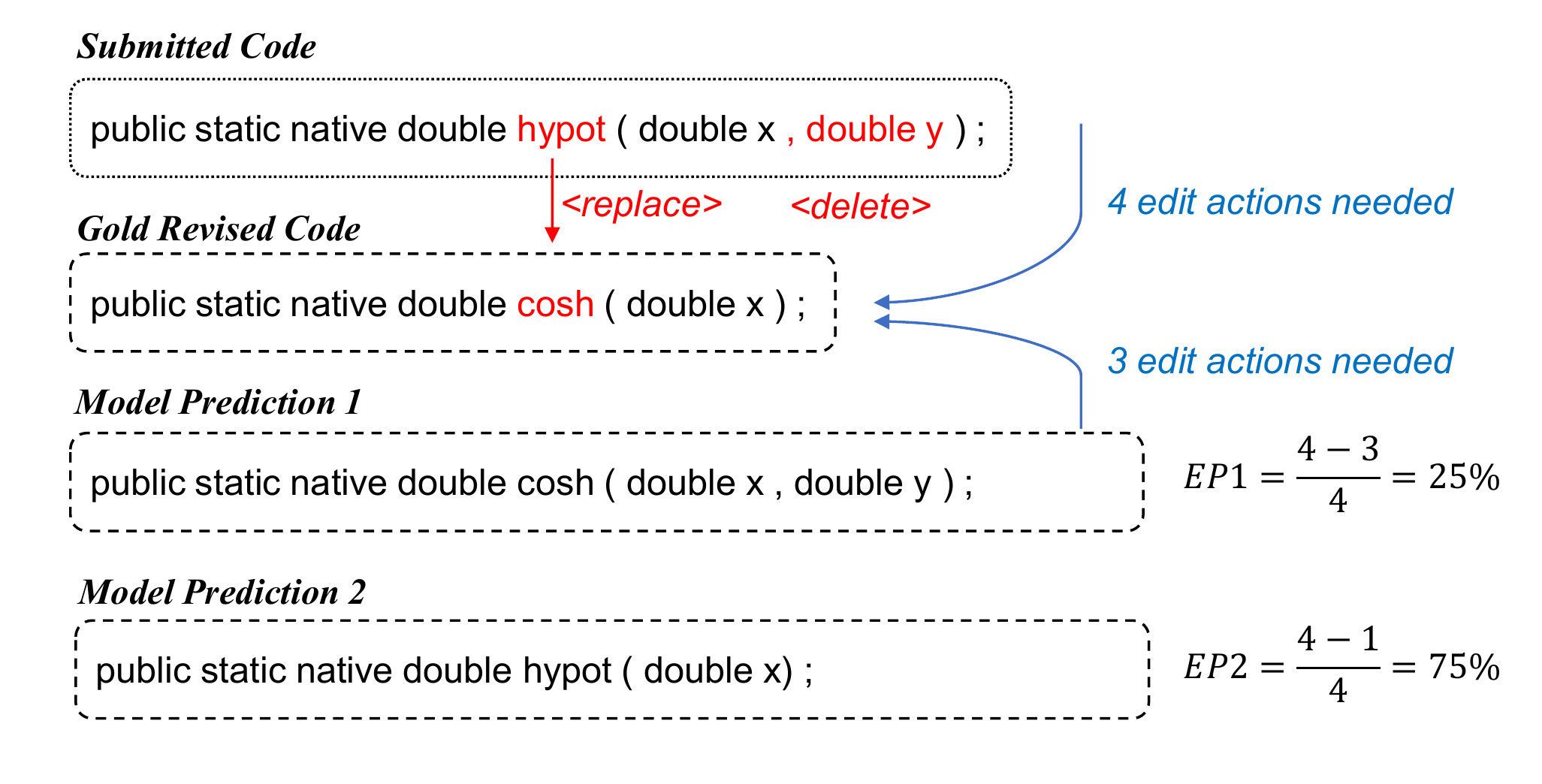}
 \vspace{-0.5cm}
	\caption{An Example of the Computation of Edit Progress Scores}
	\label{fig:edit_progress}
\end{figure}

Previous studies~\cite{svyatkovskiy2020intellicode,chowdhery2022palm,dibia2022aligning} also consider the edit distance-based similarity metric namely Normalized Edit Distance to measure the similarity between the generated code and ground truth. This metric is the character-level edit distance and can be represented as:
$Edit Distance (x, y) = \frac{Levenshtein Distance (x, y)}{max(len(x), len(y) )}$
where $x$ and $y$ are strings of the generated code and ground truth, respectively; $len(.)$ is the number of characters in the string; and $Levenshtein Distance (x, y)$ is the number of single-character edits needed to transform one string ($x$) to another ($y$).
The proposed EP is a novel metric compared to the existing ``Normalized Edit Distance'' metric:
\begin{enumerate}[leftmargin=*]
\item EP uses token-level edit distance while Normalized Edit Distance only takes into account single-character edits and may not consider the semantics of code tokens. Furthermore, single-character edits may not be suitable for pre-trained models such as CodeT5, which build their vocabularies on the token or sub-token levels and cannot directly perform single-character edits~\cite{charbert,byt5,mielke2021between}. EP uses token-level edits, which preserve the meaning of tokens and are feasible for pre-trained models.
\item EP compares the generated code with the input code to measure the generation quality of an ACR tool. In ACR tasks, the tools generate revised code, given the input. While the Normalized Edit Distance measures the similarity between the generated code and the ground truth, EP considers not only the similarity but also the improvement of the generated code compared to the input code. This makes EP more suitable than Normalized Edit Distance in situations where the input code is already very similar to the ground truth and needs to be improved further through the generated code.
\end{enumerate}

\vspace{0.1cm}
\noindent\textbf{BLEU score} is a widely used metric to measure the similarity of two natural language texts.
For review comment generation, the model generates comments written in natural language. We follow the prior work~\cite{Turian2003EvaluationOM,Coughlin2003CorrelatingAA} to adopt the commonly-used metric (i.e., the BLEU score~\cite{blue}) to measure the quality of generated texts. BLEU scores measure how well a generated text matches the ground truth text by counting the percentage of overlaps.

Please note that BLEU scores are different from Edit Progress scores: BLEU scores measure the similarity between predictions and ground truths, while Edit Progress scores measure the progress the predictions made toward generating the ground truths.

\subsubsection{\textbf{Implementation Details}}
For existing ACR tools, we adopt the original replication packages of Trans-Review\footnote{\url{https://github.com/RosaliaTufano/code_review}}, AutoTransform\footnote{\url{https://github.com/awsm-research/AutoTransform-Replication}}, and T5-Review\footnote{ \url{https://github.com/RosaliaTufano/code_review_automation}}.

By following the standard setting of fine-tuning pre-trained models~\cite{bert,codebert,GraphCodeBERT,codeT5}, we set the maximum accepted input sequence as 512, learning rate as 5e-5, and the optimizer as AdamW~\cite{AdamW}. For validation, we select the best-performing model checkpoint on the validation set. 
The best-performing checkpoint is used for testing.
We feed the test sets into each approach and get the predictions. Then, we calculate the evaluation metrics based on the predictions and the ground truths.

Both CodeBERT and GraphCodeBERT only have pre-trained encoders. Thus, when applying to a generation-based task where decoders are needed, we follow the original implementations provided by CodeBERT\footnote{ \url{https://github.com/microsoft/CodeBERT/tree/master/CodeBERT/code2nl}} and GraphCodeBERT\footnote{ \url{https://github.com/microsoft/CodeBERT/tree/master/GraphCodeBERT/translation}} in generation-based tasks to initialize a non-pre-trained decoder. 
To make the size of the decoder the same as the size of the encoder, we initialize a 12-layer Transformer decoder for CodeBERT and GraphCodeBERT to complete three generation-based code review tasks. 
We directly adopt the original CodeT5 since it has the pre-trained encode and decoder. 

Note that when generating the predictions, we allow each ACR model/pre-trained model to generate \textit{a single prediction (the Top 1 prediction)} given the input data, which is widely used in prior studies~\cite{tufano2021towards,thongtanunam2022autotransform,tufano2022using,lu2021codexglue}.

\section{Experimental Results}
\label{sec:result}
This section describes the experimental results and answers the research questions. The experimental results are summarized in Tables~\ref{tab:code_revision}, \ref{tab:comment_revision_generation}, and \ref{tab:comment_generation}, respectively. 
Note that tools in the light grey are the existing ACR tools and the others are general-purpose pre-trained models.
Regarding the performance metrics, we adopt the EM by following prior studies to investigate the effectiveness of the existing ACR tools as well as pre-trained code models in RQ1 and RQ2. 
We analyze our novel metric, EP, in RQ3 with its comprehensive rationale.

\begin{table*}[]
\caption{Experimental Results for the Code Revision Before Review (CRB) Task}
\label{tab:code_revision}
\centering
\begin{tabular}{@{}l|rrrrrr|rr@{}}
\toprule
\multirow{2}{*}{\textbf{\textbf{\begin{tabular}[c]{@{}l@{}} Approaches\\ \end{tabular}}} } & \multicolumn{2}{c}{\textbf{$\text{Trans-Review}_{\text{data}}$}} & \multicolumn{2}{c}{\textbf{$\text{AutoTransform}_{\text{data}}$}} & \multicolumn{2}{c|}{\textbf{$\text{T5-Review}_{\text{data}}$}} & \multicolumn{2}{c}{\textbf{Average}} \\ \cmidrule(l){2-3} \cmidrule(l){4-5}  \cmidrule(l){6-7} \cmidrule(l){8-9}  
                                                                                         & \textbf{EM}         & \textbf{EP}         & \textbf{EM}          & \textbf{EP}         & \textbf{EM}        & \textbf{EP}        & \textbf{EM}      & \textbf{EP}       \\ \midrule
\rowcolor[HTML]{EFEFEF} \textbf{Trans-Review}                                                                    & 3.5\%               & -1.1\%              & 0.1\%                & -16.6\%             & 0.1\%              & -151.2\%           & 1.2\%            & -56.3\%           \\
\rowcolor[HTML]{EFEFEF} \textbf{AutoTransform}                                                                   & 6.6\%               & 49.7\%              & 10.2\%               & \textbf{29.9\%}     & 0.8\%              & 9.7\%              & 5.9\%            & \textbf{29.8\%}   \\
\rowcolor[HTML]{EFEFEF} \textbf{T5-Review}                                                                       & 7.4\%               & -14.9\%             & \textbf{13.9\%}      & -71.5\%             & 3.3\%              & 13.8\%             & 8.2\%            & -24.2\%           \\
\textbf{CodeBERT}                                                                        &8.3\%                     & 49.8\%                    & 8.6\%                      & -75.3\%                     &1.1\%                   & 22.3\%                   &  6.0\%                & -1.1\%                  \\
\textbf{GraphCodeBERT}                                                                   & 6.7\%               & \textbf{50.6\%}     &  6.1\%              & -80.9\%            & 0.9\%               & 22.6\%             & 4.6\%            & -2.6\%            \\
\textbf{CodeT5}                                                                          & \textbf{8.8\%}      & 41.8\%                & 13.7\%               & -67.8\%             & \textbf{5.4\%}     & \textbf{25.6\%}    & \textbf{9.3\%}   & -0.1\%            \\ \bottomrule
\end{tabular}
\begin{flushleft}
\centering
{The gray rows refer to the existing ACR tools designed for this task and the numbers in bold indicate the best performers in each dataset and metric.}
\end{flushleft}
\end{table*}

\begin{table}[]
\caption{Experimental Results for Code Revision After Review (CRA) Task}
\label{tab:comment_revision_generation}
\centering
\resizebox{\columnwidth}{!}{%
\begin{tabular}{@{}l|rrrr|rr@{}}
\toprule
\multirow{2}{*}{\textbf{\begin{tabular}[c]{@{}l@{}} Approaches \\ \end{tabular}}} & \multicolumn{2}{c}{\textbf{$\text{Trans-Review}_{\text{data}}$}} & \multicolumn{2}{c|}{\textbf{$\text{T5-Review}_{\text{data}}$}} & \multicolumn{2}{c}{\textbf{Average}} \\ \cmidrule(l){2-3} \cmidrule(l){4-5} \cmidrule(l){6-7} 
                                                                                                  & \textbf{EM}         & \textbf{EP}         & \textbf{EM}        & \textbf{EP}        & \textbf{EM}       & \textbf{EP}      \\ \midrule
\rowcolor[HTML]{EFEFEF} \textbf{Trans-Review}                                                                             & 13.5\%              & 52.5\%              & 0.3\%              & -120.4\%                   & 6.9\%                  & -34.0\%                 \\
\rowcolor[HTML]{EFEFEF} \textbf{T5-Review}                                                                                & 24.4\%              & 65.6\%              & 9.0\%              & 38.4\%             & 16.7\%            & 52.0\%           \\
\textbf{CodeBERT}                                                                                 & 20.2\%              &57.8\%                     & 11.3\%             &     35.2\%               &  15.8\%                 &   46.5\%               \\
\textbf{GraphCodeBERT}                                                                            & 19.2\%              & 55.9\%              & 11.8\%             & 35.8\%             & 15.5\%            & 45.9\%           \\
\textbf{CodeT5}                                                                                   & \textbf{30.2\%}     & \textbf{66.8\%}     & \textbf{16.1\%}    & \textbf{41.9\%}    & \textbf{23.2\%}   & \textbf{54.4\%}  \\ \bottomrule
\end{tabular}
}
\begin{flushleft}
{The gray rows refer to the existing ACR tools designed for this task and the numbers in bold indicate the best performers in each dataset and metric.}
\end{flushleft}
\end{table}

\begin{table}[]
\caption{Experimental Results for Review Comment Generation (RCG) Task}
\label{tab:comment_generation}
\centering
\resizebox{\columnwidth}{!}{%
\begin{tabular}{@{}l|rrrr|rr@{}}
\toprule
\multirow{2}{*}{\textbf{\begin{tabular}[c]{@{}l@{}} Approaches \\\end{tabular}}} & \multicolumn{2}{c}{\textbf{$\text{Trans-Review}_{\text{data}}$}} & \multicolumn{2}{c|}{\textbf{$\text{T5-Review}_{\text{data}}$}} & \multicolumn{2}{c}{\textbf{Average}} \\ \cmidrule(l){2-3} \cmidrule(l){4-5} \cmidrule(l){6-7} 
                                                                                          & \textbf{EM}          & \textbf{BLEU}      & \textbf{EM}       & \textbf{BLEU}       & \textbf{EM}       & \textbf{BLEU}    \\ \midrule
\rowcolor[HTML]{EFEFEF} \textbf{T5-Review}                                                                        & \textbf{2.0\%}       &  \textbf{2.5\%}                  &  \textbf{2.0\%}                  & \textbf{2.0\%}        &  \textbf{2.0\%}    & \textbf{2.3\%}                 \\
\textbf{CodeBERT}                                                                         & 0.5\%                & 1.4\%                  &    0.8\%                  &  0.5\%           & 0.7\%            & 1.0\%                 \\
\textbf{GraphCodeBERT}                                                                    & 0.3\%                &  1.6\%                & 0.9\%                    & 0.6\%              & 0.6\%             & 1.1\%                 \\
\textbf{CodeT5}                                                                           & 1.1\%                &   1.4\%                 & 1.2\%                    &   1.8\%           & 1.1\%             &   1.6\%               \\ \bottomrule
\end{tabular}}
\begin{flushleft}
{The gray row refers to the existing ACR tool designed for this task and the numbers in bold indicate the best performers in each dataset and metric.}
\end{flushleft}
\end{table}

\subsection*{RQ1: Which is the best-performing ACR tool for each targeted code review activity? }

For \textbf{Code Revision Before Review} (Table~\ref{tab:code_revision}),
T5-Review achieves a performance of 7.4\%, 13.9\%, and 3.3\%  in terms of Exact Match (EM) on the three datasets, respectively. 
Among the existing ACR tools, T5-Review shows the best performance.

For \textbf{Code Revision After Review}
(Table \ref{tab:comment_revision_generation}), 
the results show that T5-Review performs the best with 24.4\% and 9.0\% EM scores in $\text{Trans-Review}_{\text{data}}$ and $\text{T5-Review}_{\text{data}}$ respectively while Trans-Review achieves 13.5\% and 0.3\% for the same datasets.

For \textbf{Review Comment Generation} (Table \ref{tab:comment_generation}), among ACR tools, only T5-Review is designed for this task. Thus, we only investigate its performance which achieves 2.0\% and 2.3\% for EM and BLEU scores on average, respectively.

Overall, we observe that T5-Review shows the best performance among state-of-the-art ACR tools on the studied downstream tasks. 
The major difference is that T5-Review is pre-trained on external large-scale data while Trans-Review and AutoTransform are not pre-trained. 
The results indicate that such pre-training is an effective solution to improve performance in most ACR tasks. 
We present further in-depth investigation on sub-findings (e.g., why Trans-Review struggles) in Section~\ref{sec:discussion}.
\begin{tcolorbox}
    \textbf{Answer to RQ1}: Among the existing ACR methods, T5-Review shows the best performance in the targeted downstream tasks, outperforming the second-best ACR tool by a large margin.
    It implies that practitioners should consider using T5-Review for ACR tasks. 
\end{tcolorbox}

 \subsection*{RQ2: Can popular pre-trained models outperform the existing ACR models?}

As we observed that pre-training may bring significant improvement in the case of T5-Review, we investigate further pre-trained code models. Although pre-training a model is usually computationally expensive~\cite{T5,codebert,codeT5},
several general-purpose pre-trained code models (i.e., CodeBERT, GraphCodeBERT, and CodeT5) are publicly released. 
In this RQ, we study the effectiveness of these pre-trained models under ACR tasks by comparing them against state-of-the-art ACR tools.

We also present the results of the general-purpose pre-trained code models in Table~\ref{tab:code_revision}, ~\ref{tab:comment_revision_generation}, and~\ref{tab:comment_generation}. 
As the observation illustrates that the pre-trained models generally outperform non-pre-trained ACR tools, we focus on the comparison between pre-trained code models and T5-Review (i.e., the best existing ACR tool).

For the \textbf{Code Revision Before Review} (Table~\ref{tab:code_revision}), 
on the one hand, CodeBERT and GraphCodeBERT fail to outperform the best ACR approach T5-Review identified in RQ1 for most cases.
Specifically, CodeBERT only outperforms T5-Review in $\text{Trans-Review}_{\text{data}}$ by 12.2\% while achieving 61.6\% and 200\% lower EM scores than T5-Review in the other two datasets.
GraphCodeBERT consistently achieves lower EM scores than T5-Review. 
On the other hand, 
CodeT5 is able to achieve higher/comparable EM scores of T5-Review. CodeT5 has significantly outperformed T5-Review in $\text{Trans-Review}_{\text{data}}$ and $\text{T5-Review}_{\text{data}}$ datasets by relative improvements of 18.9\% ($\frac{8.8-7.4}{7.4}$) and 63.6\% ($\frac{5.4-3.3}{3.3}$). However, for the $\text{AutoTransform}_{\text{data}}$ dataset, CodeT5 achieves a comparable (slightly lower) EM score than T5-Review.

For the \textbf{Code Revision After Review}
(Table~\ref{tab:comment_revision_generation}), 
CodeT5 still outperforms the best ACR tool, T5-Review, in most cases. 
Particularly, CodeT5 pushes forward the performances to new heights on EM scores of 30.2\% and 16.1\%, which are 23.8\% ($\frac{30.2-24.4}{24.4}$) and 78.9\% ($\frac{16.1-9.0}{9.0}$) over T5-Review improvements.
Moreover, CodeBERT and GraphCodeBERT get EM scores of 20.2\% and 19.2\% for the $\text{Trans-Review}_{\text{data}}$, 11.3\% and 11.8\% for the $\text{T5-Review}_{\text{data}}$, which are slightly worse than T5-Review on average.

For the \textbf{Review Comment Generation} (Table~\ref{tab:comment_generation}), CodeT5, however, is no longer the best-performing model, as T5-Review shows the best performance. 
In particular, CodeT5 only achieves 1.1\% and 1.6\% in EM and BLEU on average, which are crucially lower than T5-Review (i.e., 2.0\% and 2.3\% in EM and BLEU) on average.
Besides, both CodeBERT and GraphCodeBERT struggle with this task obtaining lower scores.

Overall, the results imply that CodeT5 is the most suitable model for Code Revision Before Review (CRB) and Code Revision After Review (CRA), while practitioners can leverage T5-Review for the Review Comment Generation. Further in-depth analysis is provided in Section~\ref{sec:discussion}.

\begin{tcolorbox}
    \textbf{Answer to RQ2}: CodeT5 has impressive performance for two of our target code review tasks (i.e., Code Revision Before Review and Code Revision After Review) by surpassing the best state-of-the-art ACR tool, T5-Review. On the other hand, T5-Review is still the best on a task (i.e., Review Comment Generation). 
    We find that pre-trained models are generally effective for code review tasks.
    
\end{tcolorbox}

\subsection*{RQ3: Comparing the generated code with the human-written code, how far apart are they?}

In this research question, we adopt the Edit Progress (EP) metric to measure the progress made by each approach on the path from the erroneous code to the golden code.
The EP score indicates the percentage of progress made. If EP equals 100\%, it means that the generated code has mitigated all the differences between the erroneous code and the ground truth. 

For the \textbf{Code Revision Before Review} task, as shown in Table~\ref{tab:code_revision}, a model that can generate the most perfect predictions (i.e., EM) may not able to generate the close-to-perfect predictions (i.e., EP).
For the $\text{Trans-Review}_{\text{data}}$ dataset, the best-performing model in EP is GraphCodeBERT, achieving an EP score of 50.6\% while CodeT5 only achieves 41.8\%.
For the $\text{AutoTransform}_{\text{data}}$ dataset, AutoTransform shows the best performance with 29.9\% in EP. Although CodeT5 performs much better than AutoTransform in EM, CodeT5 achieves much worse performance (i.e., -67.8\%) than AutoTransform. This indicates that CodeT5 generates the most perfect predictions but may generate many noisy failure cases at the same time. While AutoTransform is more conservative: it does not generate many perfect predictions but ensures that most of its revisions bring positive contributions. 
For $\text{T5-Review}_{\text{data}}$ dataset, the best model in EP is CodeT5 with a 25.6\% EP.
This indicates that we need to carefully select an approach for the CRB task when having different needs: either the one that can generate the most perfect predictions (i.e. in terms of EM) or the one that can make the most average positive contributions (i.e., in terms of EP).

Compellingly, Table \ref{tab:comment_revision_generation} reports that CodeT5 consistently outperforms all the other approaches for the \textbf{Code Revision After Review task}.
It obtains EP scores of 66.8\% and 41.9\% in both datasets while T5-Review shows comparable results (i.e., 65.6\% and 38.4\%).

Overall, the results suggest that CodeT5 is still the best model in EP for Code Revision After Review, while AutoTransform generates better code for Code Revision Before Review. Appealingly, AutoTransform achieves a 29.8\% EP score on average for Code Revision Before Review, while other approaches show negative progress, which indicates that after their revisions, the generated codes are even worse than the originally submitted code by authors on average. 
We further verify the research value of EP in Section~\ref{sec:discussion}.

\begin{tcolorbox}
    \textbf{Answer to RQ3}: 
    In summary, the EP metric can change the rank of the approaches for different code review tasks. For instance, in the Code Revision Before Review task, 
    the best model in EP (i.e., AutoTransform) is not even in the top 3 in EM.  
    This also implies there may exist much noise on the failure cases of the approaches that can generate many perfect predictions.
    CodeT5 can be the best candidate for Code Revision After Review according to the EP metric, while AutoTransform interestingly shows good performance in Code Revision Before Review for the partial progressive aspect.
\end{tcolorbox}

\section{discussion}
\label{sec:discussion}

This section presents the lessons learned from our experimental results and the threats to validity.

\subsection{Lessons Learned}

\textit{\textbf{BPE can significantly improve model performances on CRB and CRA tasks if there are new tokens introduced in ground truth codes.}}
Prior ACR tools (as well as pre-trained code models) adopt either BPE or abstraction to reduce the size of vocabulary for the ACR datasets. 
Trans-Review, the only ACR tool using abstraction, performs poorly (close to zero) on $\text{AutoTransform}_\text{data}$ and $\text{T5-Review}_\text{data}$ that contain new tokens (not shown in the submitted code) in ground truth code. 
The abstraction technique replaces actual identifiers/literals in the \textit{submitted code} with reusable IDs to form the vocabulary. However, if there are any new identifiers/literals (have not appeared in the submitted code) in the \textit{ground truth revised code}, the built vocabulary does not contain the IDs for new identifiers/literals.
Thus, it is challenging for a DL model to generate new tokens beyond the vocabulary. 
While by using Byte-Pair Encoding (BPE), all other ACR tools (e.g., CodeT5, T5-Review, and AutoTransform) show good performance on datasets with new tokens, outperforming Trans-Review by at least 700\% in EM.

\begin{table}[t]
\centering
\caption{Ratios of interrogative sentences and the model performance for RCG task.}
\resizebox{0.8\columnwidth}{!}{
\begin{tabular}{@{}lrrr@{}}
\toprule
                                                                                                                       & \textbf{CodeT5}                    & \textbf{T5-Review}                  & \textbf{Diff (\%)} \\ \midrule
\multicolumn{4}{c}{\textbf{Pretraining Data Statistics}}                                                                                                                                                                          \\ \midrule
\multicolumn{1}{l|}{\textbf{\begin{tabular}[c]{@{}l@{}}Pretraining \\ Data Source\end{tabular}}}                       & \multicolumn{1}{c|}{CodeSearchNet} & \multicolumn{1}{c|}{Stack Overflow} & -                  \\ \midrule
\multicolumn{1}{l|}{\textbf{\begin{tabular}[c]{@{}l@{}}Interrogative  \\ Sentences (\%)\end{tabular}}}                 & \multicolumn{1}{c|}{33.8\%}        & \multicolumn{1}{c|}{69.7\%}         & +106.2\%           \\ \midrule
\multicolumn{4}{c}{\textbf{Review Comment Generation ($\text{T5-Review}_{\text{data}}$)}}                                                                                                                                                \\ \midrule
\multicolumn{1}{l|}{\textbf{\begin{tabular}[c]{@{}l@{}}Subsets w/ \\ Interrogative\\ Sentences (EM)\end{tabular}}}     & \multicolumn{1}{c|}{0.24}          & \multicolumn{1}{c|}{0.84}           & +250.0\%           \\ \midrule
\multicolumn{1}{l|}{\textbf{\begin{tabular}[c]{@{}l@{}}Subset wo/ \\ Non-interrogative\\ Sentences (EM)\end{tabular}}} & \multicolumn{1}{c|}{1.60}          & \multicolumn{1}{c|}{3.02}           & +88.8\%            \\ \midrule
\end{tabular}
}
\label{tab:question_sent}
\end{table}

\textit{\textbf{The pre-trained encoder-decoder framework significantly outperforms the pre-trained encoder framework by 17.0\%--102.2\%.}}
Firstly, we find that pre-trained models perform better than not pre-trained models in most cases. For instance, pre-trained T5-Review and CodeT5 consistently show better performance than Trans-Review and AutoTransform in terms of Exact Match. 
Secondly, we find that within pre-trained models, pre-trained encoder-decoder models (e.g., CodeT5) perform significantly better (17.0\%--102.2\%) than pre-trained encoder-only models (e.g., CodeBERT and GraphCodeBERT) in generation-based code review tasks. 
The possible reason may be: when applying to generation-based tasks, pre-trained encoder-only models have to initialize a non-pre-trained decoder. This non-pre-trained decoder will lead to a worse performance~\cite{bart}. 
Lastly, within pre-trained encoder-decoder models, general-purpose pre-trained model CodeT5 shows better performance than task-specific pre-trained model T5-Review in most cases. In particular, CodeT5 could outperform T5-Review by 13.4\% to 38.9\% in terms of EM for two code revision tasks.

\textit{\textbf{Stack Overflow data is helpful to the Review Comment Generation task, especially for comments written in interrogative sentences (250.0\% improvements).}}
CodeT5 fails to outperform T5-Review for the Review Comment Generation task.
To explain why CodeT5 is inferior to T5-Review in this task, we further investigate the pre-training and downstream task datasets: 
Firstly, we find that many comments in the Review Comment Generation datasets are interrogative sentences (e.g., ``why is there a newArrayList?''). 
Then we investigate the ratios of interrogative sentences in pre-training data by a simple rule-based method: if a comment contains a question word such as  ``which", ``where", ``what", ``which", ``does", we assume this comment as an interrogative sentence.
As shown in Table~\ref{tab:question_sent}, CodeT5 is mainly pre-trained on the CodeSearchNet dataset, which contains 33.8\% interrogative sentences while T5-Review is also pre-trained on the Stack Overflow data\footnote{Because both CodeT5 and T5-Review are pre-trained on the CodeSearchNet, to study the differences between them, we focus on analyzing Stack Overflow data for T5-Review.} with a higher ratio of interrogative sentences (i.e. 69.7\% ).
Table~\ref{tab:question_sent} also presents the performances on subsets with/without interrogative sentences, using the $\text{T5-Review}_{\text{data}}$ as an example. 
We observe that with a higher ratio of interrogative sentences in the Stack Overflow data, T5-Review can outperform CodeT5 by 250.0\% on the subset with interrogative sentences.
While T5-Review only outperforms CodeT5 by 88.3\% on the subset without interrogative sentences.
The fewer interrogative sentences (comments with sentiments) in the CodeSearchNet dataset may hinder CodeT5 in generating interrogative sentences (or comments with sentiment words) given the submitted code. 

Overall, we find that Stack Overflow data is helpful to the review comment generation task (improving both with and without interrogative sentence subsets) possibly because it provides more pre-training corpus written in natural language. In particular, it can significantly improve the model's ability to generate interrogative sentences that frequently appear in human-written review comments.
This motivates future studies to consider building a large and more suitable pre-training dataset based on Stack Overflow and perform further pre-training if they aim to improve the review comment generation task.

\textit{\textbf{We advocate future work to adopt Edit Progress to measure the partial progress achieved by ACR tools.}}
The performances of code revision approaches have mostly been measured with a specific metric named Exact Match (EM).
EM is very strict as it completely ignores progressive contributions (i.e., non-perfect results) of generated code by an approach. 
To measure the partial progressive aspect, we introduced Edit Progress (EP) and evaluated the approaches with such a metric.
The overall results demonstrated that the rankings of models for each task could be changed according to which metric is being utilized.
We study and verify the research value of such a metric here. 

\begin{figure} [t]
	\centering
	\includegraphics[width=\columnwidth]{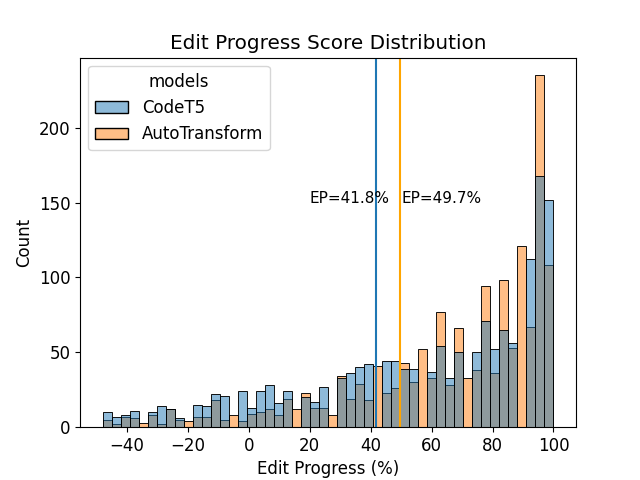}
 \vspace{-0.2cm}
	\caption{Edit Progress Scores of CodeT5 and AutoTransform in the Code Revision Before Review Task. The orange bars are the parts that AutoTransform surpasses CodeT5 in counts, and the blue bars are the parts that CodeT5 surpasses AutoTransform. The grey bars are the common parts in counts. 
 }
	\label{fig:edit_progress_score}
\end{figure}

Figure~\ref{fig:edit_progress_score} shows the EP scores of each test instance of CodeT5 and AutoTransform in the CRB task using $\text{Trans-Review}_{\text{data}}$. We choose these two approaches as they are the best-performing approaches for each metric on average in the CRB task. The vertical lines show the average EP scores of each approach.
Note that we omit the instances whose EP scores are less than -50\% for better visualization. 
As shown in Figure~\ref{fig:edit_progress_score}, CodeT5 has a better EM score because it has more generated perfect code predictions with 100\% Edit Progress scores (identical to ground truths). 
From these vertical lines (i.e., the average EP scores), we confirm that AutoTransform, which is generally inferior to CodeT5 in EM, retrieves better EP scores. 
On the one hand, AutoTransform's positive effects on the improvement of partial progress are demonstrated by the more orange-colored bars in the high EP range (e.g., 80-99\%).
On the other hand, it generates much less noise than CodeT5, as we can verify with the figure (i.e., there are much fewer orange-colored bars under 0\% EP).
This phenomenon indicates that EP can evaluate the model performances in a more comprehensive way: not only focusing on those correct predictions but also taking the failure cases into account. EP can also evaluate the practical usability of ACR tools to some extent as users do not know which prediction is correct in advance and may expect all predictions have reasonably good qualities (e.g., at least making positive contributions).

\subsection{Threats to Validity}

\newcommand{\hochkomma}{$^{,}$}

General purpose pre-trained models diverge depending on different aspects, such as the characteristics of pre-training tasks, characteristics of downstream tasks, and the size of datasets.
Our study may have a selection bias by considering several pre-trained models. We believe that employing three of the most popular pre-trained code models (i.e., CodeBERT, GraphCodeBERT, and CodeT5) can mitigate the threat.
Another threat to validity regards the dataset selection, as it may deliver bias in the experimental results. In our work, we have minimized this threat by using multiple datasets, i.e., $\text{Trans-Review}_{\text{data}}$~\cite{tufano2021towards}, $\text{AutoTransform}_{\text{data}}$~\cite{thongtanunam2022autotransform}, and $\text{T5-Review}_{\text{data}}$~\cite{tufano2022using}.
Furthermore, we publicly share the replication package\footnote{Replication Package. \url{https://github.com/soarsmu/Generated_Review_How_Far}} including datasets for future comparisons by the research community.
The evaluation metrics used may also bring a selection bias.
To reduce it, we reuse the same evaluation metric (i.e., Exact Match) with existing ACR tools~\cite{tufano2021towards, thongtanunam2022autotransform, tufano2022using} and the widely used metric for texts (i.e., BLEU~\cite{blue}).
Besides, to study the partial progress achieved by ACR tools, we follow prior work in text editing~\cite{elgohary-etal-2021-nl} to adopt the Edit Progress metric. 

\section{related work}
\label{sec:relate}

In this section, we review the research work that relates to our work: approaches to automating code review activities and pre-trained models for SE.

\noindent\textbf{Automating Code Review Activities.} 
We focus on three generation-based code review automation tasks: Code Revision Before Review~\cite{tufano2021towards,thongtanunam2022autotransform,tufano2022using}, Code Revision After Review~\cite{tufano2021towards,tufano2022using}, and Review Comment Generation~\cite{tufano2022using}.
Other studies focus on similar code review activities in different manners.  
Hellendoorn et al.~\cite{hellendoorn2021towards} aimed to predict the locations of code change hunks that possibly need review and revision. 
They showed that locating a suitable hunk is a challenging task.
Siow et al.~\cite{siow2020core} proposed CORE, a multi-level embedding approach to represent the semantics of code changes and reviews, which recommends code reviews in a retrieval-based way.
Hong et al.~\cite{hong2022commentfinder} proposed a retrieval-based approach to recommend code review comments. 
Li et al.~\cite{li2022automating} developed CodeReviewer, a pre-trained model for code review. They evaluated it on three tasks using new datasets, aiming to improve code review. In contrast, we aimed to compare existing ACR tools and identify future directions.

In addition, there are other tasks to automate other code review activities to support software developers, such as reviewer recommendation~\cite{thongtanunam2015should,rahman2016correct,sulun2019reviewer,al2020workload}, review prioritization~\cite{fan2018early,shi2019automatic,li2019deepreview,wu2022turn} and defect-proneness prediction in submitted code~\cite{kamei2012large,mcintosh2018fix,deepjit, cc2vec,pornprasit2021jitline,zhou2022simple}.
For instance, Thongtanunam et al.~\cite{thongtanunam2015should} found 4\% to 30\% of reviews in open-source projects could not find suitable reviewers (i.e., reviewer recommendation). 
Rahman et al.~\cite{rahman2016correct} proposed CORRECT which utilizes external library similarity and technology expertise similarity of reviewers to recommend reviewers. 
Al-Zubaidi et al.~\cite{al2020workload} did not only focus on reviewer experience but also take into account the review workload when recommending reviewers.

\section{Conclusion and Future Work}
\label{sec:conclusion}

In this paper, we empirically study the existing generation-based automatic code review (ACR) techniques and the general-purpose pre-trained code models. 
We evaluate the effectiveness of these tools on three representative generation-based code review tasks.
Our experimental results show that CodeT5 is the best performer in most cases, outperforming the best existing ACR tool T5-Review by 13.4\%--38.9\% in Code Revision Before/After Review tasks.
However, in Review Comment Generation, T5-Review outperforms all other methods by a large margin.
Our work also has revealed several interesting findings that can motivate future work.

In the future, we plan to design a new ACR tool that can perform well both in Exact Match and Edit Progress metrics.

\vspace{0.1cm}
\noindent{\bf Acknowledgement.} This research / project is supported by the National Research Foundation, Singapore, under its Industry Alignment Fund – Pre-positioning (IAF-PP) Funding Initiative. Any opinions, findings and conclusions or recommendations expressed in this material are those of the author(s) and do not reflect the views of National Research Foundation, Singapore.

\balance
\bibliographystyle{IEEEtran}
\bibliography{sample}

\end{document}